\documentclass[preprint,showpacs]{revtex4}
\begin{document}

\title{Pion Source in Au+Au
Collisions at $\sqrt{s_{_{NN}}}=130$ GeV}

\author{Bi Pin-zhen}
\affiliation{Institute of Modern Physics, Fudan University, Shanghai 200433, China}
\email{pzbi@fudan.edu.cn}

\date{\today}

\begin{abstract}
The modified model emission function S(x,p), with a uniform radial expansion mode and  mass reduction, 
is used to study the single particle spectra and two particle correlation of pions. 
The calculated HBT radii and one particle spectra are compared 
with the data from $\sqrt{s_{_{NN}}}=130$ GeV Au+Au collisions at the
Relativistic Heavy Ion Collider (RHIC).  We see the main features of the data are reproduced,
$R_{o}/R_{s}$ decreases with increasing $K_{T}$ and we have a  kinetic freeze out temperature $T = 150 MeV$. 
\end{abstract}

\maketitle

\section{ Introduction}

We believe quarks are constituents of hadrons, but at present we have not found "free quarks" yet.  At low densities quarks are
confined in individual hadrons, we expected that at high energy density nuclear matter undergoes a transition to deconfined 
"quark matter" or "quark gluon plasma", where quarks could move "freely" in a volume much larger than that of a single hadron.
The primary goal of the ultrarelativistic heavy ion collision is to create the new form of matter,
quark gluon plasma. Bose-Einstein correlation in multiparticle production processes provide valuable
information on the space-time dynamics of the fundamental interactions. It is expected that correlations
of identical pions produced in the ultrarelativistic heavy ion collisions could lead to a better understanding  
of the properties of the new matter form. 
The main aim of particle interferometric methods is to extract as much information as possible about the emission 
function $S(x,K)$, which characterizes the particle emitting source created in the heavy ion collision.
A completely model independent reconstruction of emission function from measured correlation data is not possible,
only certain combinations of spatial and temporal source characteristics are measurable \cite{Hei}, and
some  useful pieces of information about the phase-space distribution $S(x,K)$ could be obtained . One particle spectra
and two particle HBT (Hanbury-Brown-Twiss) correlation functions are related to it in \cite{Shuryak,Pratt,Chapman}
\begin{eqnarray}
E{{dN} \over {d^{3}p}} &=& {\int{d^{4}xS(x,p)}},                       \\
C(K,q)&=&1+{{|\int{d^{4}xS(x,K)}e^{iq{\cdot}x}|^{2}}  \over {\int{d^{4}xS(x,p_{1})}\int{d^{4}xS(x,p_{2})}}}       
\end{eqnarray}
where $K=(p_{1}+p_{2})/2$ ,$q=(p_{1}-p_{2})$ and $p_{1}$, $p_{2}$ are momentums of detected particles.
Experimental measurements of C are usually parameterized in terms of the intercept  ${\lambda}(K)$ and the HBT radii 
$R_{ij}^{2}(K)$ by
\begin{eqnarray}
C(K,q)=1+{\lambda} \ {\exp{[-{{\sum}}  R^{2}_{ij}(K)q_{i}q_{j}]}}.
\end{eqnarray}
The relative momentum is decomposed into 
components parallel to the beam (l=longitudinal), 
parallel to the transverse component
of K (o=out), and in the remaining third direction (s=side).
A basic model for $S(x,K)$ is proposed in ref.\cite{Akk,Hei}. We expect that its main characteristics can be quantified 
by its widths in the spatial and temporal directions, and a collective dynamical component.

\begin{eqnarray}
S(x,p)&=&{{2J+1} \over {{(2{\pi})^{3}}{\pi}{\Delta}{\tau}}}{m_{\bot}}{\cosh (y-{\eta})}
\exp{[- {{p{\cdot}u(x) - {\mu}}  \over {T}}]}       \nonumber        \\
&    &{\times}\exp{[ - {{{r}^{2}}  \over  {2R^{2}}} 
- {{{\eta}^{2}} \over  {2({\Delta}{\eta})^{2}}}
- {{({\tau}-{\tau}_{0})^{2}} \over {2({\Delta}{\tau})^{2}}}]}.
\end{eqnarray}
Here ${\tau}=\sqrt{t^{2}-z^{2}}$ denotes the longitudinal proper time and ${\eta}={{1} \over {2}}{\ln}[(t+z)/(t-z)]$
the space-time rapidity.
The transverse mass $m_{\bot}={\sqrt{p^{2}_{\bot}+ m^{2}}}$.
Spin degeneracy $(2J+1) = 1$  ,  and chemical potential ${\mu}= 0$ for pion.
The particle four-momentum can be expressed by the momentum rapidity and the transverse mass,

\begin{eqnarray}
p_{\mu}   &=& (m_{\bot}{\cosh{y}}, p_{\bot}, 0, m_{\bot}{\sinh{y}}),   \\
{{p{\cdot}u(x)} \over {T}} &=& {{m_{\bot}} \over {T}}{\cosh{(y-{\eta})}}{\cosh{{\eta}_t}}- {{p_{\bot}} \over {T}}{{x} \over {r}} {\sinh{{\eta}_{t}}}.
\end{eqnarray}
The source has a finite geometrical size in the spatial and temporal directions encoded in 
transverse and longitudinal Gaussian widths R and ${\Delta}{\eta}$ as well as in a finite 
particle emission duration ${\Delta}{\tau}$.
The freeze out temperature are mainly determined by the 
the transverse single pion spectra  \cite{Bearden} and resonance decay contributions 
are often taken into account in fitting processes \cite{Heinz2} . The influence of collective velocity is considered also with
a linear transverse flow rapidity profile 
\begin{eqnarray}
{\eta }_{t}(r)={\eta}_{f}{{r} \over {R}}
\end{eqnarray}
in the model \cite{Heinz3}.
With these assumptions the HBT radii can be calculated by
\begin{eqnarray}
R^{2}_{s}(K)=<{\tilde{y}}^{2}>(K),      \\
R^{2}_{o}(K)=<({\tilde{x}} - {\beta}_{\bot}{\tilde{t}})^{2}>(K),       \\
R^{2}_{l}(K)=<({\tilde{z}} - {\beta}_{l}{\tilde{t}})^{2}>(K)
\end{eqnarray}
where $<...>$ denotes an average with the emission function $S(x,K)$:
\begin{eqnarray}
<f>(K) = {{\int{d^{4}xf(x)S(x,K)}}   \over  {\int{d^{4}xS(x,K)}}}.
\end{eqnarray}
The space-time coordinates ${\tilde{x}}^{\mu}$ are defined relative to the "effective source center"  ${\bar{x}}^{\mu}$ by
\begin{eqnarray}
{\tilde{x}}^{\mu}(K)={x^{\mu}}-{\bar{x}}^{\mu}(K),     \nonumber     \\
{\bar{x}}^{\mu}(K)= <x^{\mu}>(K).
\end{eqnarray}
But at present it is difficult  to reproduce one particle spectra and the HBT radii extracted from two-pion correlations simultaneously.
$R_{o}/R_{s}$ from hydrodynamics is greater than one, and increases as $K_{T}$ does.  In stead, experiment finds that 
$R_{o}/R_{s}$ decreases with increasing $K_{T}$ \cite{Pisar}. Different approaches are proposed to deal 
with this problem recently \cite{Bron,Csor,Ster} . Here
we would like to modify this basic model and try to find a set of parameters, by which the main features of the one 
particle spectrum and HBT radii could be reproduced.

\section{The modified model}

(1)We use Bose-Einstein distribution function instead of Boltzmann function. The difference is not small for pion at low transverse momentum.
The resonance decay contributions are not included. Only the direct pions are considered,
\begin{eqnarray}
S(x,p)&=&{{2J+1} \over {{(2{\pi})^{3}}{\pi}{\Delta}{\tau}}}{m_{\bot}}{\cosh (y-{\eta})}         \nonumber     \\ 
&    &{\times}{{1} \over {\exp{[{{p{\cdot}u(x) - {\mu}}  \over {T}}]} - 1}}   \nonumber      \\
&    &{\times}       
\exp{[ - {{{r}^{2}}  \over  {2R^{2}}} 
- {{{\eta}^{2}} \over  {2({\Delta}{\eta})^{2}}}
- {{({\tau}-{\tau}_{0})^{2}} \over {2({\Delta}{\tau})^{2}}}]}
\end{eqnarray}

(2)The radial expansion mode is assumed to be uniform,  independent of radial position, ${\eta}_{t}=C$.
We are motivated to do this by the fact that the experiment shows a type of "explosive" behavior, which implies
that a strong pressure is build-up at the origin. We first generalize the linear transverse flow rapidity profile  to 
\begin{eqnarray}
{\eta}_{t}(r)={\eta}_{f}({{r} \over {R}})^{\alpha}.
\end{eqnarray}
It is easy to see,  a smaller $\alpha$ implies a stronger radial flow. Then we take ${\alpha}=0$ to express the strongest  radial
flow, the strongest pressure build-up at the origin. Maybe a ${\delta}$ function could be used to describe the pressure at the origin.
We neglect  the detailed structure in this small region near the center
and express it in a simplified way. 

(3)In the procedures of fitting, the temperature influence on the pion mass is considered.
Many different models are used to approach this problem, their results are quite different.
We only use the effective pion mass as an adjustable parameter here, its value is fitted by the data.
We will compare it with some model estimated values in section III.

Under these modifications,  a set of parameters are found, $ m_{\pi}=40 MeV$, $T=150 MeV$,  ${\eta}_{t}=0.55$, $R=5.2fm$,  ${\Delta}{\eta}=5.6$,
${\tau}_{0}=6 fm/c$,  and ${\Delta}{\tau}=1fm/c$.   

The $P_{T}$ dependence of the yield of ${\pi}^{-}$ is fitted by
\begin{eqnarray}
{{1} \over {P_{T}}}{{d^{2}N} \over {dP_{T}dy}}}=A{\int{d^{4}xS(x,P)}
\end{eqnarray}
It is assumed that all particles decouple kinematically at the same freeze out  temperature.  
The experimental data are taken from ref.\cite{Adcox}. Transverse momentum spectra of ${\pi}^{-}$
are measured  at midrapidity 
for the most central events. The results from the emission function with mass reduction, uniform radial expansion  and Bose-Einstein
distribution
are shown in fig.1 by the solid line. It is fitted in a broad range $P_{T} < 2.2 GeV$.  The dashed line shows the results
from the emission function with the uniform radial expansion and Bose-Einstein distribution,
but without mass reduction.  We see that the influence of mass reduction
exists in low transverse momentum region.  The dot dashed line shows the results from the emission function with linear radial expansion
and without mass reduction (other parameters are the same as given above) .  The curve is changed greatly. 
This implies that the extracted temperature is connected with the 
expansion mode tightly.


In fig.2 solid line shows the calculated results of $R_{s}$  by the emission function with mass reduction and uniform radial expansion.
Data of $R_{s}$ for pions as a function of $K_{T}$ are measured at mid-rapidity 
from $\sqrt{s_{_{NN}}}=130$ GeV Au+Au collisions at 
RHIC. 
Solid circle (empty circle)  for ${\pi}^{+}$ (${\pi}^{-}$) are given by ref.\cite{Adcox2}.
Triangle-down (triangle-up) for ${\pi}^{+}$ (${\pi}^{-}$) are given by ref.\cite{Adler}.
We see the main features of the data are reproduced by the modified emission function.
The dot dashed line shows the results from the emission function without modifications.


The experimental data for $R_{o}$ and $R_{l}$ are also compared with the calculated results  in fig.3 and fig.4 respectively.
The results from the emission function with 
mass reduction and uniform radial expansion  are shown by the solid lines 
and the results from the emission function without modifications are shown by the dot dashed lines.
We see the slope of $R_{o}$ is larger in the modified model. 
The ratio ${R_{o}}/{R_{s}}$ obtained from the emission functions is shown in fig.5 by the solid line (with modifications) and
the dot dashed line (without modifications).
Data are from PHENIX  \cite{Adcox2}
and STAR \cite{Adler}. 
The pion source parameters given here do not reproduce the data exactly  yet,  but it is consistent with the main features of the
single particle spectra and HBT radii, $R_{o}/R_{s}$ decreases with increasing $K_{T}$.
If this picture is acceptable, we have a  kinetic freeze out temperature $T = 150 MeV$. It is 
just near the chemical freeze out temperature.  


\section{Discussions}
(1)We take the uniform radial expansion mode here. It does not mean that there is only pressure (and pressure gradients) at the
origin of the system. It only means that pressure gradients at the small region near the origin of the system is much stronger than
that at the other parts. It is only a simplified description.  We take this choice as it can produce a reasonable result and it is consistent
with the picture of explosion. 

(2)The kinetic freeze out temperature is mainly determined by the one particle spectra.  With different expansion mode the extracted
temperature is different.  In fig.1 the dashed line shows the result from the emission function with linear expansion profile and without 
mass reduction. It is only used to show the influence of the modification. 
Our result $T=150 MeV$ is just near the chemical freeze out temperature. It implies that the chemical freeze
out and kinetic freeze out maybe occur simultaneously.

(3)A basic idea for deconfinement is that the vacuum can have two possible states. The first is the normal vacuum outside the hadrons.
Quarks and their gluon fields modify the vacuum in their neighborhood, transforming it into the second state,  the "perturbative" vacuum,
or "bag" state,  the vacuum inside the hadron. The inside state of the vacuum has a higher energy per unit volume than the outside
vacuum state,  the energy difference , B, is often called as bag constant in the bag model. The current quark mass is about 10 MeV for
u quark and d quark. The mass of hadron with light quarks, i.e. the energy of a static spherical bag with radius R, is
$E={{C} \over {R}}+BR$ where V is the volume of the bag and C is a positive constant, which is determined by the number
of quarks in the bag and the eigenvalues of energy and zero-point energy \cite{Aerts,Koch}. Determined by $dE/dR=0$, the energy of the bag 
has a minimum value $E={4/3}{(4{\pi}B)}^{1/4}C^{1/4}$. R.D.Pisarski gives the temperature dependent bag constant
$B(T){\propto}{(T^{4}_{c}}-{T^{4})}$ in \cite{Pisa}. The field theory at finite temperature  also shows the bag constant is temperature 
dependent \cite{Bi1,Shi},  $B(T)={B_{0}}{(1-{{T^2} \over {T^2_c}})^2}$  where $T_c$ is the critical temperature.
At the critical temperature the difference between the normal vacuum and the perturbative vacuum disappears, and the quarks are
deconfinement. Near the critical temperature the difference between the two vacuums is small and thus the mass of hadron
is reduced. At the freeze out condition, the hadrons decouple from each other, 
the mass of a hadron really has its vacuum value, but as the temperature around it
is still high, it has a "hot vacuum"  value.  A "free hadron" in the vacuum is not "alone", the vacuum around it is full of virtual particles.
For hadrons with heavy quarks, H. Satz  use potential at finite temperature to study their properties \cite{Satz1}.
The detailed expressions for mass shift are different in these approaches, but they have common points. The hadron mass decreases with the 
increasing temperature \cite{Bi5}. 
The mass shift is also studied in other ways. R. D. Pisarski  and F. Wilczek \cite{Pisa2} study the restoration of the spontaneously broken
chiral symmetry in a linear $\sigma$ model and show that pion mass is stable but under some conditions the mass of ${\pi}_0$
could drop to 77 MeV. Pion is taken as Goldstone bosen its mass would be zero, but in the realistic case, pion mass has a small value.
G. E. Brown and M. Rho studied the chiral restoration in hot and/or dense matter, and showed that the explicit chiral
symmetry breaking--that  gives masses to the Goldstone bosons could be "rotated away" in \cite{Brown}.
We also use linear $\sigma$ model to study the properties of $\pi^{0}$ \cite{Bi3}, we find that if the explicit breaking term is temperature
independent pion mass could even increase as temperature does, but the temperature dependent coupling will change the situation.
From the above discussions we see the predictions from different models are quite different. Only the experiments could 
give judges. Unfortunately we have not found enough evidence \cite{Bi6} to show that the hadron mass (and  pion mass) decreases with
increasing temperature at present.  It is difficult to measure the hot hadron mass directly as the observer is in low temperature
environment, the pion mass returns to the normal value there.
From the view point of bag model, the volume of pion shrinks and obtains the energy from the vacuum (bag constant), 
thus the pion mass returns to the normal value. In this process, the inter structure of pion is changed. 
Such a change will give no influence on the $P_T$ distribution of pion.  
We could measure the influence of mass shift , but in some cases the influence 
could be even washed away by other facts.
In our fitting procedures, we only take pion mass as an adjustable parameter to fit the data.
Our result seems consistent with the mass reduction picture and  previous studies \cite{Tak1,Tak2,Has,Bi2, Bi4}.

(4)The correlation function is determined by the momentum distributin at the freeze out temperature in both
the original model and the modified model. The difference of the momentum distributions at the freeze out 
temperature between these two models are caued different thermal distributions before the freeze out.

(5)A completely model-independent reconstruction of the emission function S(x,p) is not possible.  We follow the approach given 
in refs.\cite{Hei,Akk}
by using simple parametrizations of the emission function.  We adjust the parameters by a comparison to data.  The modified model
can reproduce the main
features of the one particle spectra and HBT radii simultaneously now. This model should be further improved, if it is used 
to describe the azimuthal momentum space  anisotropy of particle emission \cite{Kolb}.  
The pion production from resonance is not considered here. In ref.\cite{Hei}, resonance decay contributions are shown.
Thus a detailed calculation including these effects is necessary for us to make the model better.
We will discuss it elsewhere.


\newpage
Captions:

Fig.1 Transverse momentum spectra measured at mid-rapidity for ${\pi}^{-}$ at most centrality 
given in ref.\cite{Adcox} are compared with results obtained from the emission function with mass reduction, uniform
radial expansion and Bose-Einstein distribution(solid line), with uniform radial expansion and Bose-Einstein distribution, but
without mass reduction (dashed line), and  with linear
radial expansion and Bose-Einstein distribution but without mass reduction (dot dashed line).

Fig.2 $R_{s}$ for pions as a function of $K_{T}$ are measured at mid-rapidity.
Solid circle (empty circle)  for ${\pi}^{+}$ (${\pi}^{-}$) are given by ref.\cite{Adcox2}.
Triangle-down (triangle-up) for ${\pi}^{+}$ (${\pi}^{-}$) are given by ref.\cite{Adler}.
Solid line shows the calculated results from the emission function with mass reduction and uniform radial expansion.
Dot dashed line shows the results from the emission function without modifications.

Fig.3 $R_{o}$ for pions as a function of $K_{T}$ are measured at mid-rapidity.
Solid circle (empty circle)  for ${\pi}^{+}$ (${\pi}^{-}$) are given by ref.\cite{Adcox2}.
Triangle-down (triangle-up) for ${\pi}^{+}$ (${\pi}^{-}$) are given by ref.\cite{Adler}.
Solid line shows the calculated result from the emission function with mass reduction and uniform radial expansion.
Dot dashed line shows the results from the emission function without modifications.

Fig.4 $R_{l}$ for pions as a function of $K_{T}$ are measured at mid-rapidity.
Solid circle (empty circle)  for ${\pi}^{+}$ (${\pi}^{-}$) are given by ref.\cite{Adcox2}.
Triangle-down (triangle-up) for ${\pi}^{+}$ (${\pi}^{-}$) are given by ref.\cite{Adler}.
Solid line shows the calculated result from  emission function with mass reduction and uniform radial expansion.
Dot dashed line shows the results from the emission function without modifications.

Fig.5 ${R_{0}}/{R_{s}}$ for pions as a function of $K_{T}$ are measured at mid-rapidity.
Solid circle (empty circle)  for ${\pi}^{+}$ (${\pi}^{-}$) are given by ref.\cite{Adcox2}.
Triangle-down (triangle-up) for ${\pi}^{+}$ (${\pi}^{-}$) are given by ref.\cite{Adler}.
Solid line shows the calculated result from  emission function with mass reduction and uniform radial expansion.
Dot dashed line shows the results from the emission function without modifications.

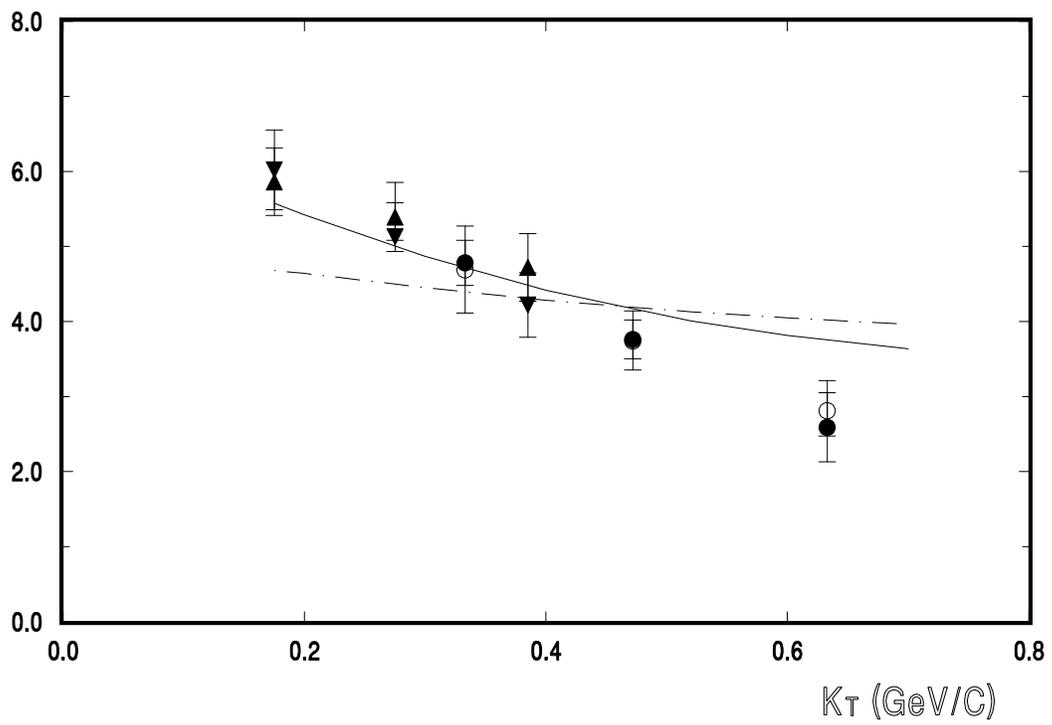
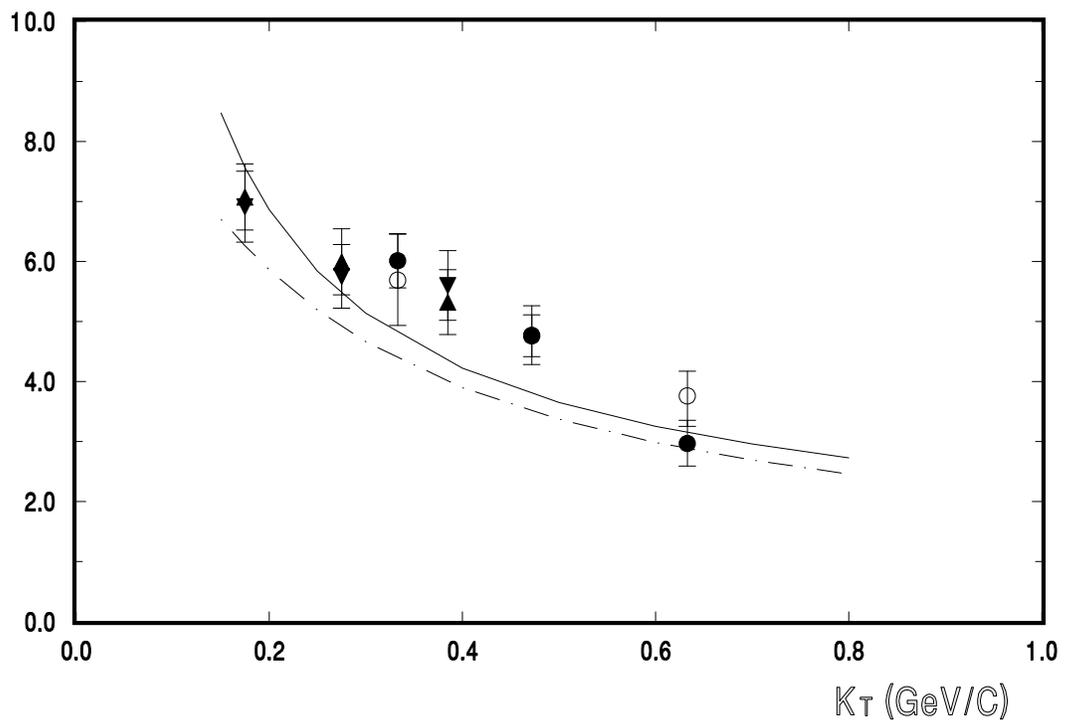
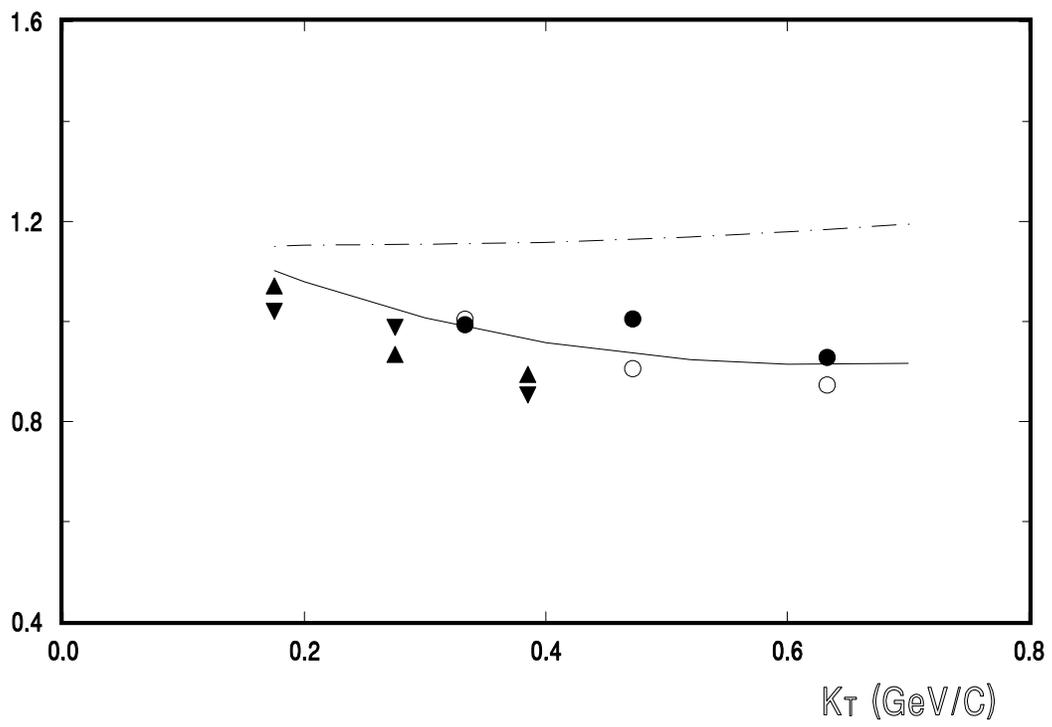
\end{document}